\documentclass[useAMS,usenatbib,epsfig, a4paper,amsmath,aps]{mn2e}
 
\usepackage{ulem}
\usepackage{float}
\usepackage{epsfig}
\usepackage{subfigure}
\usepackage{epstopdf}
\usepackage{verbatim}
\usepackage{nicefrac}
\usepackage[draft]{hyperref}
\usepackage{url}
\usepackage{mathrsfs}
\usepackage{amssymb}
\usepackage{amsmath}
\usepackage{bm}
\usepackage{wasysym}
\usepackage{ulem}
\usepackage{dsfont}
\usepackage{upgreek}

\title[Velocities \& Lensing on the Hubble diagram]
   {The effects of velocities and lensing on moments of the Hubble diagram}
\author[E.~Macaulay et al.]
{E.~Macaulay$^{1,2,3}$\thanks{email: \href{mailto:e.macaulay@port.ac.uk}{\nolinkurl{edward.macaulay@port.ac.uk}}}, T. M. Davis$^{1,2}$,  D. Scovacricchi$^3$, D. Bacon$^3$, T. Collett$^3$,  R. C. Nichol$^3$\\ 
$^{1}$School of Mathematics \& Physics, The University of Queensland, St. Lucia, 4072, Brisbane, Australia\\
$^{2}$ARC Centre of Excellence for All-sky Astrophysics (CAASTRO), Sydney, 2006, Australia\\
$^{3}$Institute of Cosmology and Gravitation, University of Portsmouth, Burnaby Rd, Portsmouth, PO1 3FX, UK
}

\date{\today}
\pubyear{2016}
\begin{document}
\maketitle
\begin{abstract}

We consider the dispersion on the supernova distance-redshift relation due to peculiar velocities and gravitational lensing, and the sensitivity of these effects to the amplitude of the matter power spectrum.  We use the MeMo lensing likelihood  developed by Quartin et al., which accounts for the characteristic non-Gaussian distribution caused by lensing magnification with measurements of the first four central moments of the distribution of magnitudes.  We build on the MeMo likelihood by including the effects of peculiar velocities directly into the model for the moments.  In order to measure the moments from sparse numbers of supernovae, we take a new approach using Kernel Density Estimation to estimate the underlying probability density function of the magnitude residuals.  We also describe a bootstrap re-sampling approach to estimate the data covariance matrix.  We then apply the method to the Joint Light-curve Analysis (JLA) supernova catalogue.  When we impose only that the intrinsic dispersion in magnitudes is independent of redshift, we find $\sigma_8=0.44^{+0.63}_{-0.44}$ at the one standard deviation level, although we note that in tests on simulations, this model tends to overestimate the magnitude of the intrinsic dispersion, and underestimate $\sigma_8$.  We note that the degeneracy between intrinsic dispersion and the effects of $\sigma_8$ is more pronounced when lensing and velocity effects are considered simultaneously, due to a cancellation of redshift dependence when both effects are included.  Keeping the model of the intrinsic dispersion fixed as a Gaussian distribution of width 0.14 mag, we find $\sigma_8 = 1.07^{+0.50}_{-0.76}$.  

  \end{abstract}


\section{Introduction}
\label{sec:intro}

Understanding the nature of dark energy is one of the key goals of modern cosmology.  A recurring theme in understanding dark energy is measuring both the growth of structures and geometry of the Universe \citep[e.g.][]{2008MNRAS.389L..47A,2010PhRvD..81j3510Z,2012PhRvD..85l3546Z,2013MNRAS.429.1514S,2014MNRAS.439.3504S,2015PhRvD..91f3009R}.  In this work, we consider using supernovae, typically considered a probe of geometry, to constrain the growth of structure.

\subsection{Growth and Geometry}

Here, geometry refers to observational probes that are predominantly sensitive to the background expansion of the Universe: observations that constrain the distance-redshift relation.  Among geometrical probes, observations are either `standard rulers', such as Baryon Acoustic Oscillations \citep[e.g.][]{2003ApJ...598..720S,2003ApJ...594..665B,2005ApJ...633..560E,2011MNRAS.418.1707B,2012MNRAS.427.3435A,2013A&A...552A..96B,2014MNRAS.441...24A}, or `standard candles', such as supernovae Ia \citep[e.g.][]{2009ApJ...700..331H,2009ApJS..185...32K,2010A&A...523A...7G,2010AJ....139..120F,2011ApJS..192....1C,2014A&A...568A..22B,2016arXiv160303823N}.  Supernovae are well known as a key observable probe in establishing the accelerated expansion of the Universe \citep{1998AJ....116.1009R,1999ApJ...517..565P}.  These geometrical observables constrain the densities of the constituents of the Universe (such as the matter density, $\Omega_{\rm{m}}$), and the equations of state of these densities.  A key aim of these observations is to measure $w$, the equation of state of dark energy.

With the accelerating expansion of the Universe now well established \citep[e.g.][]{2008ARA&A..46..385F,2015arXiv150201590P}, and ever more precise measurements of $w$, new dark energy observations are focusing on measuring the growth of the Universe \citep[e.g.][]{2005PhRvD..72d3529L,2012MNRAS.425..405B}.  Here, growth refers to the growth of cosmological density perturbations.  The motivation for measuring the growth of density perturbations is that theoretical models of dark energy (which must reproduce the observed distance-redshift relation) often predict different expectations for the growth of density perturbations \citep[e.g.][]{2012PhR...513....1C}. 

Among growth probes, there is a natural division of the observations into two kinds: either relativistic or non-relativistic.  Relativistic techniques such as gravitational lensing \citep{2001PhR...340..291B,2012MNRAS.427..146H} and the integrated Sachs Wolfe effect \citep{1967ApJ...147...73S,2003astro.ph..7335S,2005PhRvD..72j3519P,2006PhRvD..74f3520G,2015arXiv150201595P} are sensitive to the path that photons take, and are therefore sensitive to both time-like and space-like perturbations to the metric.  On the other hand, non-relativistic observations, such as Redshift Space Distortions, \citep{1987MNRAS.227....1K,2001Natur.410..169P,2008Natur.451..541G,2012MNRAS.426.2719R} focus on the positions and velocities of large scale structure.  These observations are sensitive only to time-like perturbations to the metric, the `Newtonian' potential.

A key feature of general relativity (GR) is that time-like and space-like perturbations to the metric should be equal \citep[e.g.][]{2007PhRvD..76j4043H,2013PhRvD..87b4015B}, and so measuring both Newtonian and Lensing density fluctuations is a powerful technique to constrain modified gravity alternatives to a cosmological constant dark energy \citep[e.g.][]{2010Natur.464..256R,2013MNRAS.429.2249S,2013A&A...557A..54D,2015JCAP...12..051L,2015arXiv151104457P,2016MNRAS.456.2806B}. The equality of these two potentials is only true in the absence of anisotropic stress, as we would expect for cold dark matter (CDM), so an inequality may also suggest dark matter interactions.  We thus have several observational tests that any theory of the dark Universe must pass: the distance-redshift relation, and measurements of Newtonian and Lensing density fluctuations.  

Supernovae are one of the most established probes of the distance-redshift relation.  In this paper, we focus on using supernovae to also constrain Newtonian and Lensing density fluctuations.  The main motivation for this work is to contribute to the development of a new method to measure the amplitude of the matter power spectrum with supernovae.  $\sigma_8$ is the amplitude of the matter power spectrum at scales of 8 $h^{-1}$Mpc, and a key parameter to measure in order to constrain cosmological density fluctuations, although as we will see, supernova magnitude fluctuations are sensitive to fluctuations on very different physical scales.

The best-fitting value for $\sigma_8$ from the Planck measurements of the cosmic microwave background is 0.830$\pm$0.015 \citep{2015arXiv150201589P}.  However, it is important to remember that this is a derived value (not a measurement) that depends on the measured value of the amplitude of primordial density fluctuations ($A_s$), and a cosmological model.  Current measurements of $\sigma_8$ from weak lensing shear and redshift space distortions find a value that is lower than expectations from the $\Lambda$CDM model with Planck parameters \citep[e.g.][]{2013MNRAS.429.1514S,2013PhRvL.111p1301M,2014MNRAS.444..476R,2016PDU....13...66C,2015arXiv150201590P,2016arXiv160600439G}.  Depending on the choice of model (such as imposing flatness, or a cosmological constant), the tension is at the level between two and three standard deviations.

While the tension remains only moderate, and unaccounted-for systematic effects in the observables cannot be ruled out, there is the possibility that the tension may represent some of the first hints of physics beyond the $\Lambda$CDM model.  New methods to measure $\sigma_8$ are important to determine if the current tension is due to new physics, or systematic effects.

\subsection{Signals in the Noise}

A more general motivation for this work is to develop a method to measure cosmological parameters from signals that are often considered as noise.  \cite{2013JCAP...06..002B,2013PhRvL.110b1301B} modelled the effect of cosmological density fluctuations on luminosity distance measurements.  They found that peculiar velocities and gravitational lensing are the dominant sources of the dispersion, and place a fundamental limit on our ability to measure parameters affecting the distance redshift relation, such as $\Omega_{\Lambda}$.  In this work, instead of considering the dispersion from lensing and peculiar velocities as noise in the distance-redshift relation, cosmological signals in this distribution are our primary focus.

At low redshift, peculiar velocities can be inferred for individual galaxies with an independent distance indicator, such as the Tully-Fisher, Fundamental Plane or supernovae \citep[e.g.][]{2012ApJ...751L..30H,2012MNRAS.423.3430B,2016MNRAS.455..386S}.  This can be achieved at higher redshifts if galaxy sizes or magnitudes are cross-correlated with galaxy overdensities \citep{2014MNRAS.443.1900B}. Also at higher redshift, peculiar velocities can be inferred statistically in a galaxy redshift survey, via redshift space distortions \citep[e.g.][]{2011MNRAS.415.2876B,2015arXiv151108083O,2016PhRvD..93b3525S}.  

\cite{2007PhRvL..99h1301G} analysed the peculiar velocities of 271 low redshift supernovae, and found $\sigma_8=0.79 \pm 0.22$.  \cite{2015JCAP...12..033H} analysed the bulk flow of the 100 lowest redshift galaxies in the joint light-curve analysis (JLA) catalogue.  Instead of fitting for $\sigma_8$ directly, the analysis fitted the amplitude $A$ of the peculiar velocity covariance matrix, normalized to the expected value in $\Lambda$CDM.  The analysis found a value of $A$ consistent with zero peculiar velocities, but with a large uncertainty that also included the expected $\Lambda$CDM value.

The magnitudes of standard candles are also affected by gravitational lensing.  In rare cases, when the photons pass close to a massive cluster, a supernova can be strongly lensed, or even lensed into multiple images \citep[e.g.][]{2015Sci...347.1123K}.  However, all supernovae will be weakly-lensed by cosmological density perturbations.  \cite{2005ApJ...633..589S} detected lensing magnification in the cross-correlation of quasar and foreground galaxy correlations.  One of the most established techniques for detecting weak gravitational lensing is by measuring coherent distortions in the shapes of galaxies \citep[e.g.][]{2014MNRAS.441.2725F,2015arXiv150705552T}.  In contrast, supernova lensing is sensitive to the change in magnitude caused by weak gravitational lensing. Since lensing depends on the integrated path of the photons, the effect is most significant for high redshift supernovae \citep[e.g.][]{2016MNRAS.456.1700S}.

\cite{2006ApJ...640..417G} considered correcting for the lensing dispersion in the Hubble diagram by estimating the magnification effect from large scale structure, and \cite{2010MNRAS.402..526J} considered using supernovae magnification to constrain the properties of the lensing dark matter haloes.  \cite{2013MNRAS.432..679C} considered modelling the line of sight lensing signal to improve time delay measurements.  \cite{2014ApJ...780...24S} tested for lensing magnification with 608 supernovae from the Sloan Digital Sky Survey by cross-correlating the magnitude residuals with the expected lensing signal from foreground galaxies.  Although the significance of the cross-correlation was low at the 1.4 standard deviation confidence level, the correlation suggests that lensing provides some contribution to the distances measured with supernovae.

\cite{2006PhRvD..74f3515D} proposed using the dispersion in the Hubble diagram due to weak lensing to measure $\sigma_8$, but noted that the Gaussian model used in their work can be biased towards incorrect values of $\sigma_8$ due to the non-Gaussian nature of the lensing dispersion.  In a series of papers, \cite{2013PhRvD..88f3004M} and \cite{2014PhRvD..89b3009Q} developed a method to measure $\sigma_8$ from the effect of lensing magnification on the magnitude residuals of supernovae (called Method-of-the-Moments -- MeMo).  In \cite{2014MNRAS.443L...6C}, the MeMo technique was applied to the JLA and Union2 supernovae catalogues, finding $\sigma_8=0.84^{+0.28}_{-0.65}$ at the 68\% confidence level, or $\sigma_8 < 1.45$ at the 95\% confidence level.  

In \cite{2016PDU....13...66C}, the MeMo lensing likelihood was combined with a peculiar velocity likelihood. These two different physical effects were combined by using a peculiar velocity likelihood for supernovae with redshift $z<0.1$, and the lensing-only MeMo likelihood for higher redshifts.  This approach has the advantage that correlations between supernovae from large scale bulk flows can be modeled in the velocity likelihood, but does not model the combined effect of the two different kinds of perturbations on the moments.  \cite{2016PDU....13...66C} allowed both $\sigma_8$ and the perturbation growth index $\gamma$ to vary, finding the best-fitting $\sigma_8=0.65^{+0.23}_{-0.37}$, with $\gamma=1.35^{+1.7}_{-0.65}$.  Keeping $\gamma$ fixed at the expected value in GR of 0.55, the best-fitting value of  $\sigma_8$ was $0.40^{+0.21}_{-0.23}$.

Our approach to treating lensing and velocities is different from that of \cite{2016PDU....13...66C}.  Instead of treating the two effects as independent -- with two different likelihoods -- we use a single likelihood that directly combines predictions for lensing and velocities into the expectations for the moments.  The advantage of this approach is that the total expectations for the moments include contributions for both effects, which would otherwise be underestimated.

\cite{2006PhRvD..74f3515D} assumed that the intrinsic dispersion in supernova magnitudes can be modeled with a Gaussian distribution with mean zero and a standard deviation that is independent of redshift.  \cite{2014MNRAS.443L...6C} relaxed this assumption somewhat by allowing the intrinsic dispersion to be further modeled with intrinsic third and fourth moments (although also constant in redshift).  In both papers, the only variation in the distribution of residuals was assumed to be due to perturbations in the metric, as a function of $\sigma_8$.  In reality, the intrinsic dispersion in the magnitudes of supernovae may vary with redshift, and may not be Gaussian.  For example, Malmquist bias may affect the distribution of fainter residuals, sub-populations of different types of Ia supernovae may skew the intrinsic dispersion, or correlations with host-galaxy evolution may introduce redshift dependence \citep[e.g.][]{2009ApJ...691..661H,2010ApJ...722..566L,2010MNRAS.406..782S,2010ApJ...715..743K,2016MNRAS.457.3470C}.  \cite{2016arXiv160408885S} tested the Hubble residuals of recent supernova data with the Kolmogorov Smirnov test, and found the residuals to be consistent with a Gaussian distribution.  \cite{2014MNRAS.443L...6C} tested models of the intrinsic dispersion that are both constant and vary with redshift, and found that the Bayesian evidence favoured a model for the intrinsic dispersion that is constant in redshift.

In this work, we aim to place limits on cosmological density fluctuations via the effects of peculiar velocities and lensing magnification on moments of the Hubble diagram.  We verify the MeMo lensing likelihood on simulated catalogues from the MICE light cone simulation.  We include the effects of peculiar velocities directly into the moments likelihood, as opposed to including velocities with a separate likelihood.  In Section \ref{sec:method_modelling} we describe the physical model of the moments, based on the effects of peculiar velocities and lensing magnification. In Section \ref{sec:data} we review the JLA catalogue and the simulated realizations of the catalogue that we generate.  In Section \ref{sec:measurements} we describe our techniques to measure the moments and estimate the data covariance matrix.   In Section \ref{sec:results} we present the results of fitting for the lensing and velocity models to the JLA catalogue, and compare our results to other analyses.  In Section \ref{sec:conclusions}, we summarize our main conclusions.

\section{modelling the Effects of Structure on Supernova Magnitudes}
\label{sec:method_modelling}

Throughout this paper we assume a flat $\Lambda$CDM model.  We consider the perturbed Friedmann-–Lema\^itre–-Robertson–-Walker metric, given by
\begin{equation}
\label{eq:metric}
ds^2 = -a^2(1+2\Psi)d\eta^2 + a^2 (1- 2\Phi) {\boldsymbol{dx}} ^2,
\end{equation}
where $\Psi$ represents time-like perturbations to the metric, and $\Phi$ represents space-like perturbations.   In the case of CDM and GR, the two potentials are equal, and so any departure from 
$\Psi = \Phi$ may suggest physics beyond $\Lambda$CDM.  Traditionally, supernovae have been used to probe only the history of the scale factor, $a$, by measuring the distance modulus, $\mu$ (in Mpc)
\begin{equation}
\label{eq:modulus}
\mu = 25 + 5 \log_{10}\left( D_{\rm{L}} \right),
\end{equation}
where, for a flat universe, the luminosity distance $D_{\rm{L}}$ is given by
\begin{equation}
\label{eq:luminosity_distance}
D_{\rm{L}} = (1+z_{\rm{obs}})  \int{ dz \frac{c}{H(z)} }
\end{equation}
where $z$ is the cosmological redshift (i.e., without a peculiar velocity), and $z_{\rm{obs}}$ is the redshift including the additional Doppler shift due to the peculiar velocity.  In order to fully constrain the physics of the dark Universe, we must also constrain perturbations to the metric.  Both peculiar velocities and lensing constrain perturbations, and depend on the amplitude of the matter power spectrum, characterized by $\sigma_8$.  Peculiar velocities are sensitive to very large scale modes in the matter power spectrum, while lensing is sensitive to small scale modes.  Our assumption in this work is that the amplitude of the power spectrum can be calculated for all $k$ scales once one has set $\sigma_8$.  Any variation in the signals from lensing and velocity may be due to scale dependent effects in the matter power spectrum.  For the small $k$ scales to which lensing is sensitive, the matter power spectrum is also sensitive to complicated baryonic physics \citep{2006ApJ...640L.119J,2010MNRAS.405.2161D,2011MNRAS.417.2020S}.

\subsection{Peculiar Velocities}

At low redshift, deviations from the Hubble flow are dominated by peculiar velocities.  The peculiar velocity covariance matrix is given by \citep[e.g.][]{2015JCAP...12..033H}
\begin{equation}
\label{eq:PV_CM}
\left< \mu_i, \mu_j \right> = \left( \frac{5}{\ln 10} \right)^2 \left( \frac{(1+z_i)^2}{H(z_i)D_{\rm{L}}(z_i)} \right) 
\left( \frac{(1+z_j)^2}{H(z_j)D_{\rm{L}}(z_j)} \right) \xi,
\end{equation}
where the velocity correlation function $\xi$ can be decomposed into components that are parallel and perpendicular to the line of sight, following the notation in \cite{2007PhRvL..99h1301G},
\begin{equation}
\label{eq:PV_para_perp}
\xi = \sin \theta_i \sin \theta_j \xi_{ \perp} + \cos \theta_i \cos \theta_j \xi_{||}.
\end{equation}
The parallel and perpendicular components are given by
\begin{equation}
\label{eq:PV_corFunc}
\xi_{||,\perp}= H_0^2 f(z_i)  f(z_j) a_i a_j \int{ \frac{dk}{2 \uppi^2} P(k) K_{||,\perp}(kr)  },
\end{equation}
where the window functions $K$ are
\begin{equation}
\label{eq:K_para}
K_{||}(x) = j_0(x) - \frac{2 j_1(x)}{x}
\end{equation}
and
\begin{equation}
\label{eq:K_perp}
K_{\perp}(x) = \frac{ j_1(x)}{x}.
\end{equation}
$ j_0(x)$ and $ j_1(x)$ are Bessel functions.  For $i=j$ (i.e., the diagonal of the covariance matrix), the arguments of the window functions $K$ are $x=0$.  Since $K_{\perp}$ tends to 0 as $x$ tends to 0, and $K_{||}$ tends to $\frac{1 }{3}$ \citep{1988ApJ...332L...7G}, the diagonal of the covariance matrix is given by (in km s$^{-1}$)
\begin{equation}
\label{eq:PV_coeff}
\sigma_{\rm{V,kms}^{-1}}^2 = H_0^2 f^2(z) a^2  \int{ dk \frac{ P(k) }{2 \uppi^2}  \frac{1 }{3}},
\end{equation}
which we can convert into magnitudes:
\begin{equation}
\label{eq:PV_coeff_mags}
 \sigma_{\rm{V}} = \frac{5}{\ln 10} \left( \frac{(1+z)^2}{H(z)D_{\rm{L}}(z)} \right) \sigma_{\rm{V,kms}^{-1}} .
\end{equation}
The form of equation (\ref{eq:PV_coeff_mags}) can also be used to include contributions from non-linear peculiar velocities (which are not modeled by the power spectrum)
\begin{equation}
\label{eq:PV_coeff_NL}
 \sigma_{*} = \frac{5}{\ln 10} \left( \frac{(1+z)^2}{H(z)D_{\rm{L}}(z)} \right) \sigma_{\rm{V,NL}},
\end{equation}
where $\sigma_{\rm{V,NL}}$ is the non-linear velocity dispersion, in km s$^{-1}$.  We can include the non-linear dispersion in the likelihood as an additional parameter to marginalize over, although we find that this has only a minimal effect on the results.

\subsection{Lensing}

At higher redshifts ($z \gtrsim 0.4$), the effects of lensing magnification dominate the effects due to peculiar velocities.  The change in magnitude $\Delta \mu$ due to weak lensing magnification is related to the lensing convergence $\kappa$ along the line of sight by \citep[e.g.][]{2001PhR...340..291B}
\begin{equation}
\label{eq:mag}
\Delta \mu = 5 \log_{10}(1-\kappa),
\end{equation}
or, to first order in $\kappa$
\begin{equation}
\label{eq:mag}
\Delta \mu \approx -\frac{5}{\ln 10} \kappa .
\end{equation}
$\kappa$ is given by a  weighted sum of the density fluctuations $\delta$ along the line of sight,
\begin{equation}
\label{eq:kappa_theory}
\kappa = \frac{3 \Omega_{\rm{m}} H_0^2}{2c^2} \int_0^{\chi_S}{ d \chi  \left[ (\chi_S - \chi) \frac{\chi}{\chi_S} \left(1+z_{\chi} \right) \right] \delta(\chi) }.
\end{equation}
The term in square brackets weights the contribution of the density fluctuation by the fractional distance of $\chi$ to the source, $\chi_S$, at redshift $z_{\chi}$.  The dispersion in  $\kappa$ is given by \citep[e.g.][]{2014MNRAS.443.1900B}
\begin{equation}
\label{eq:kappa_disp_theory}
\bar{\kappa}^2 = \frac{9 \Omega^2_{\rm{m}} H_0^4}{4c^4} \int_0^{\chi_S}{ d \chi  \left[ (\chi_S - \chi) \frac{\chi}{\chi_S} \left(1+z_{\chi} \right) \right]^2 \int{dk  \frac{kP(k)}{2 \uppi} } }.
\end{equation}
Combining equation (\ref{eq:mag}) and (\ref{eq:kappa_disp_theory}), we can relate the power spectrum to the expected lensing dispersion in magnitudes:
\begin{equation}
\label{eq:mag_kappa}
\sigma^2_{\rm{L}} \approx \left( \frac{5}{\ln 10} \right)^2 \bar{\kappa}^2
\end{equation}
We do not use equations (\ref{eq:mag_kappa}) and (\ref{eq:kappa_disp_theory}) directly in our likelihood.  Instead, we use a fitting function given by equation 6 in \cite{2013PhRvD..88f3004M}, which we have verified reproduces the theoretical expectation.

However, the dispersion due to lensing magnification has a characteristic, negatively skewed non-Gaussian distribution with a long magnification tail.  The large tail of magnified lines of sight is due to the rare, densest lines of sight, which cause the photons along these paths to be more focused, leading to a magnification.  Conversely, most lines of sight in the Universe are under dense.  Photons travelling along these paths are consequently de-magnified (compared to a path at average density).

\subsection{Moments of the Residuals}

In order to measure this characteristic non-Gaussianity, the approach we take here is to build on the MeMo lensing likelihood presented in \cite{2014PhRvD..89b3009Q}.  The MeMo approach fitted for the effects of lensing magnification on the first four moments of the Hubble residuals.  These moments are related to the variance, skewness and kurtosis of the residuals.  The model in \cite{2014PhRvD..89b3009Q} did not include the effects of peculiar velocities, and the combined lensing and velocity analysis in \cite{2016PDU....13...66C} used two independent likelihoods for the lensing and velocities.  However, the two effects cannot be split into two independent likelihoods, since the additional dispersion caused by the peculiar velocities contributes to the moments measured in the lensing likelihood.

Instead, our approach is to include the additional dispersion from peculiar velocities directly into the moments analysis.  The advantage of this approach is that the additional dispersion (that is greater than the lensing effect for $z<0.4$) can be directly accounted for in the moments analysis.  The disadvantage is that correlations in magnitudes cannot be included -- we are effectively modelling only the diagonal component of the distance modulus covariance matrix.  However, we find that the full velocity covariance matrix given by equation (\ref{eq:PV_CM}) is dominated by the diagonal component.  For $z>0.1$, the correlations between the supernovae are almost entirely negligible.  Moreover, this approach allows us to correctly calculate the total effect on the moments due to lensing and velocities.

The $i$th moment $\mu_i$ of a variable $\mu$ is defined as
\begin{equation}
\label{eq:moments_def}
\mu_i \equiv  \left<    (   \mu - \left< \mu \right>   )^i \right> ,
\end{equation}
or, equivalently,
\begin{equation}
\label{eq:moments_def_integral}
\mu_i =  \int{  d\mu (   \mu - \left< \mu \right>   )^i } P(\mu),
\end{equation}
where $\left< \mu \right>$ is the mean of the distribution and $P(\mu)$ is the probability distribution of $\mu$.  The MeMo likelihood is given by
\begin{equation}
\label{eq:MeMO_like}
L = \exp \left(  -\frac{1}{2} \sum_j \chi_j^2 \right),
\end{equation}
where $j$ is a sum over redshift bins.  Within each redshift bin, the $\chi^2$ is given by
\begin{equation}
\label{eq:MeMO_chiSq}
\chi^2 = ( \boldsymbol{\mu} - \boldsymbol{\mu_{\rm{fid}}})^t \mathsf{C} ^{-1} ( \boldsymbol{\mu} - \boldsymbol{\mu_{\rm{fid}}}),
\end{equation}
where $\mathsf{C}$ is the data covariance matrix in the $j$th redshift bin. 

We use the same redshift binning as \cite{2014MNRAS.443L...6C}, with equally spaced bins of $\Delta_z=0.1$, to a maximum of $z=0.9$, and find that the results are not significantly affected by the redshift binning.  We use the ensemble sampler \textsc{emcee} \citep{2013PASP..125..306F} to probe the posterior distribution of the parameters.  Since the covariance matrix is estimated from the data, and does not depend on the parameters that are fitted, equation (\ref{eq:MeMO_like}) does not depend on the determinant of the covariance matrix.  We describe our method to estimate the covariance matrix in Section \ref{sec:cov_mat}.

\begin{figure*}
\begin{center}
 \includegraphics[width=18cm]{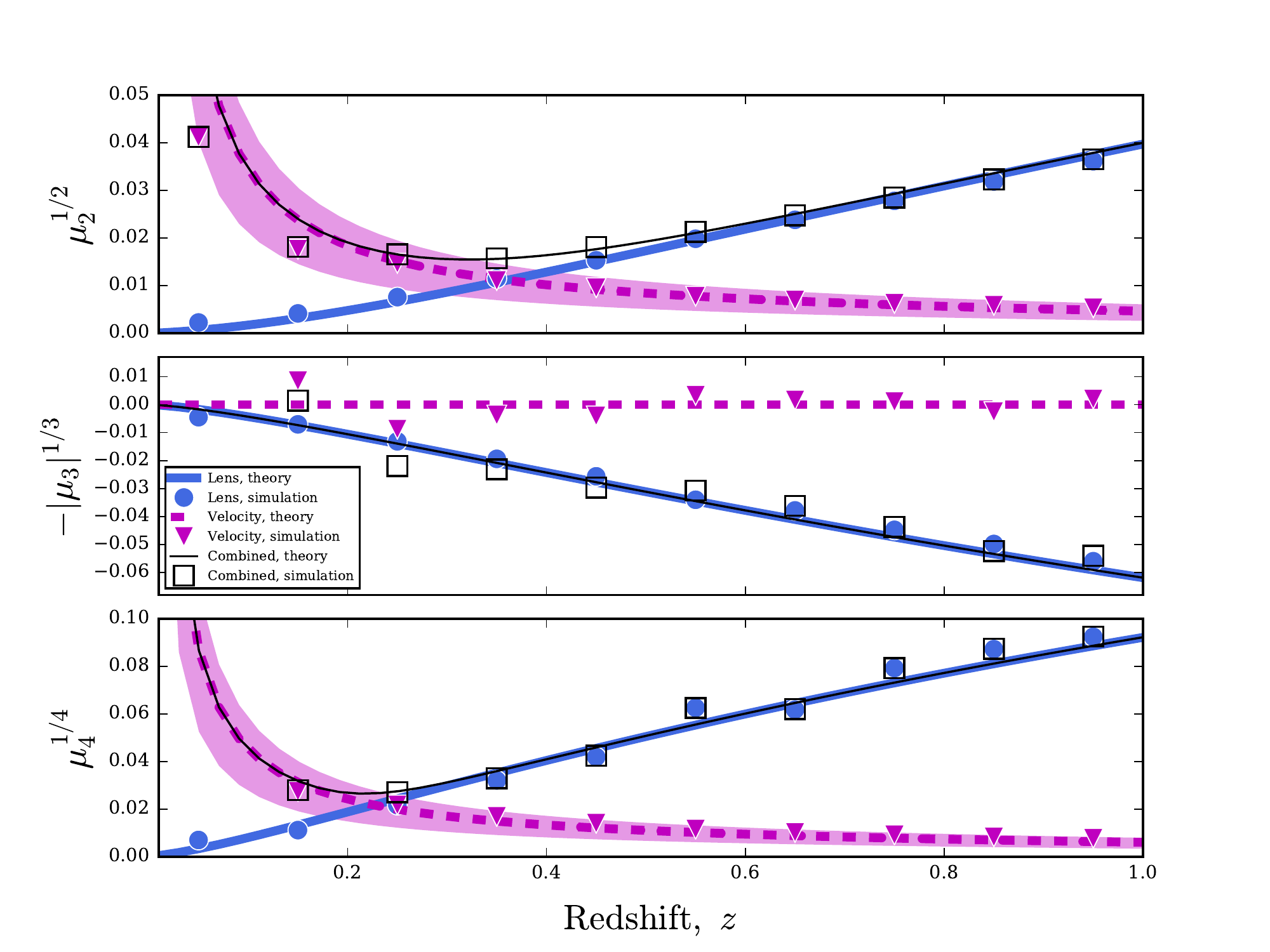}
\caption{Comparing theoretical predictions for the moments of the distribution of magnitude residuals due to velocities and lensing.  The lines show the analytical predictions, and the points show values from a simulated lightcone.  The dispersion due to peculiar velocities increases towards $z=0$, as the peculiar velocities grow, and contribute a larger fraction to the observed redshift, as illustrated by the dashed line.  For the velocity model, we also illustrate with the shaded region the effect of non-linear velocity dispersion.  The lower edge of the band corresponds to zero velocity dispersion, increasing to 500 km s$^{\rm{-1}}$ at the top of the band.  The velocity model plotted includes a non-linear dispersion of 350 km s$^{\rm{-1}}$.  The dispersion due to lensing magnification increases linearly with redshift, as the photons travel greater distances over which they can be dispersed.  Due to the shape of the distribution, lensing also adds a characteristic skewness and kurtosis to the distribution, which both also increase with redshift.}
   \label{fig:moments_theory}
\end{center}
\end{figure*}

Here, $\boldsymbol{\mu}$ is a vector of the observed first four moments of the distance moduli within the redshift bin
\begin{equation}
\label{eq:MeMO_muVec}
\boldsymbol{\mu} = \{ \mu_1, \mu_2, \mu_3, \mu_4 \},
\end{equation}
and $\boldsymbol{\mu_{\rm{fid}}}$ is the corresponding vector of theoretical expectations for the moments.  The second moment is given by
\begin{equation}
\label{eq:mu2_combine}
\mu_2 =  \sigma^2_{\rm{I}} + \sigma^2_{\rm{L}} + \sigma^2_{\rm{V}},
\end{equation}
where $\sigma_{\rm{L}}$ is the lensing dispersion, $\sigma_{\rm{V}}$ is the velocity dispersion, and $\sigma_{\rm{I}}$ is the intrinsic dispersion in supernova magnitudes. The non-linear velocity dispersion $\sigma_{*}$ can also be added in quadrature to this expression, although we find that including this parameter has only a small effect on the results.  We assume that $\sigma_{\rm{I}}$ is constant, and independent of redshift.  The third moment is given by
\begin{equation}
\label{eq:mu3_combine}
\mu_3 =  \mu_{3,\rm{L}} + \mu_{3,\rm{I}} ,
\end{equation}
where  $\mu_{3,\rm{L}}$ is the contribution due to lensing, and we also allow an intrinsic $\mu_{3,\rm{I}}$ to vary.  Due to isotropy, we would expect large scale structure to cause peculiar velocities moving equally towards or away from the line of sight.  As such we would not expect peculiar velocities to preferentially magnify or de-magnify, so we do not include a contribution to the third moment from velocities.  We verify that velocities do not contribute to the skewness of the residuals in Fig. \ref{fig:moments_theory}. The fourth moment is given by
\begin{equation}
\label{eq:mu4_combine}
\mu_4 =  \mu_{4,\rm{L}} + \mu_{4,\rm{I}} + 3 \mu^2_2 - 3 \sigma^4_{\rm{L}},
\end{equation}
where, as before, $\mu_{4,\rm{L}}$ are the intrinsic and lensing contributions to the moment $\mu_{4,\rm{I}}$.  We subtract the $3 \sigma^4_{\rm{L}}$ term so that the equation reduces to equation 11 in \cite{2014PhRvD..89b3009Q} in the absence of additional contributions from velocities.  The effect of the velocities affects the moment via the $ 3 \mu^2_2$ term.  We calculate $\sigma_{\rm{V}}$ and $\sigma_{*}$ from equations (\ref{eq:PV_coeff_NL}) and (\ref{eq:PV_coeff_mags}).  For the lensing moments, we use fitting functions given by equations 6, 7 and 8 in \cite{2013PhRvD..88f3004M}, which have been calibrated to $N$-body simulations for a range of cosmological parameters.

For $\mu_1$, we use equation (\ref{eq:modulus}) to calculate the expected distance modulus from the observed supernovae.  In Fig. \ref{fig:moments_theory} we compare the moments of the distribution of magnitude residuals.  We note that the model for the second moment agrees with the theoretical modelling from \cite{2013JCAP...06..002B}.

\section{Supernova Data \& Simulations}
\label{sec:data}

\begin{figure}
\begin{center}
 \includegraphics[width=9cm]{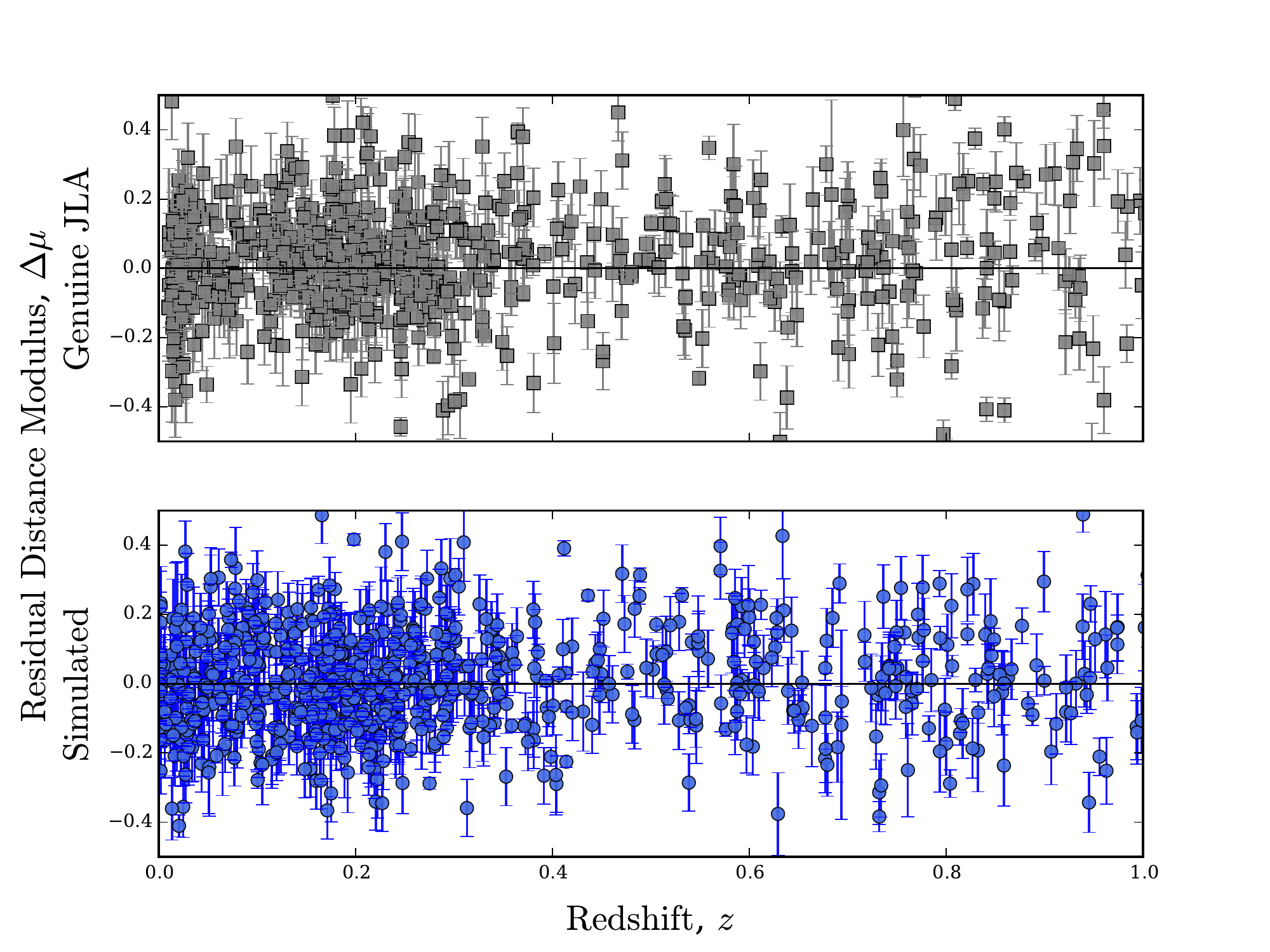}
\caption{Hubble diagram residuals.  The top panel shows the genuine JLA survey, the lower panel shows one of the simulated realizations.}
   \label{fig:residuals_JLA_sim}
\end{center}
\end{figure}

Throughout this paper, we consider distances measured with the JLA supernova catalogue \citep{2014A&A...568A..22B}.  We calculate the observed distance modulus using
\begin{equation}
\label{eq:mu_obs}
\mu = m_B^{\star} - (M_B - \alpha X_1 + \beta C),
 \end{equation}
where $X_1$ is the stretch parameter of the light curve and $C$ is the colour parameter. $m_B^{\star}$ is the observed $B$-band peak magnitude.  We also apply a stellar mass correction,
\begin{equation}
\label{eq:mass}
M_B = M_B^1 + \Delta_M,
 \end{equation}
for $M_{\rm{stellar}}>10^{10} M_{\astrosun}$.  We calculate the distance modulus for the best-fitting values from \cite{2014A&A...568A..22B} of $\alpha$, $\beta$, $M_B$ and $\Delta_M $, given by $\alpha=0.141$, $\beta=3.101$, $M_B=-19.05$, and $\Delta_M=-0.07$.  The residuals of these distance moduli (after a best-fitting cosmology has been subtracted) are shown in Fig. \ref{fig:residuals_JLA_sim}.  The error bars are the square-root of the diagonal of the covariance matrix.  We use these uncertainties to weight our estimates of the central moments.

The publicly available redshifts from \cite{2014A&A...568A..22B} are the observed heliocentric redshift, $z_{\rm{hel}}$, and the heliocentric corrected, CMB rest frame redshift, $z_{\rm{cmb}}$.  However, the $z_{\rm{cmb}}$ have also been adjusted to subtract the effect of peculiar velocities, which have been estimated from local density fields.  For our analysis, these peculiar velocities are a signal, not a nuisance.  To recover the heliocentric corrected redshifts without the additional peculiar velocity corrections, we take the $z_{\rm{hel}}$ and subtract only the heliocentric correction.

\subsection{Simulated Catalogues}
\label{sim_section}

To test our analysis, we generate simulated realizations of the JLA catalogue by sampling galaxies from the MICE cosmological light cone simulation \citep{2015MNRAS.448.2987F}, with cosmological parameters $\Omega_{\rm{m}}=0.25$, $\Omega_{\Lambda}=0.75$, $H_0=70.0$ kms$^{-1}$ and $\sigma_8=0.8$. The MICE simulation includes peculiar velocities, and an estimate of the lensing convergence, but does not include effects of baryonic physics on the evolution of the density perturbations.

To generate a simulated realization of the JLA, we start with a subsample of the MICE simulation with 8.9 million galaxies in the redshift range 0 to 1.4.  We then find the galaxy in the MICE catalogue that has the closest redshift to each supernovae in the JLA.  For each of these galaxies, we save the cosmological redshift $z_{\rm{cos}}$, the redshift with the additional contribution from peculiar velocities, $z_{\rm{pec}}$ and the lensing convergence $\kappa$ values.  We first calculate the unlensed distance modulus with $z_{\rm{cos}}$, $z_{\rm{pec}}$ and equations (\ref{eq:modulus}) and (\ref{eq:luminosity_distance}).  We then use the $\kappa$ value and equation (\ref{eq:mag}) to calculate the magnification effect on the distance modulus.  Next, we add in an intrinsic dispersion sampled from a normal distribution.  We choose a value of 0.14 mag for the intrinsic dispersion, to match the value expected in the JLA.  We then add together the distance modulus, lensing effect, and intrinsic dispersion to generate the `observed' distance modulus.  We assign the quoted uncertainty from the genuine supernova to the simulated counterpart.  We use the value of $z_{\rm{pec}}$ from the MICE simulation as the `observed' redshift.

To generate multiple independent simulated realizations, we repeat the procedure as before for each realization, but instead match the simulated redshift to the genuine redshift shifted by a value sampled from a normal distribution with mean zero and standard deviation of 0.03.  The value of 0.03 was chosen to ensure that each realization would be independent, without affecting the overall redshift distribution.  We use this procedure to generate 10 realizations of the JLA catalogue.  10 realizations are sufficient to assess the variation in the simulated catalogues while generating further catalogues is limited by computational resources required to evaluate the full likelihood.

\section{Measuring the Moments}
\label{sec:measurements}

In this section, we consider methods for measuring moments, and specifically focus on the accuracy and precision of the measurements from sparse data. For $N$ points of discrete data, 
\begin{equation}
\label{eq:moments_def_discrete}
\mu_i= \frac{\sum   \left( \Delta_{\mu}  \right)^i}{N},
\end{equation}
which is often used to estimate the moment, where $\Delta_{\mu}=  (   \mu - \left< \mu \right>   )$.  We can also weight this estimate by the uncertainty in  $\mu$, $\sigma_{\mu}$:
\begin{equation}
\label{eq:moments_def_discrete_weight}
\mu_i= \frac{\sum  \sigma^{-2}_{\mu} \left( \Delta_{\mu}  \right)^i}{\sum \sigma^{-2}_{\mu}}.
\end{equation}
However, this estimate only converges to an unbiased estimate of the moment in the limit of large $N$, particularly so for distributions with large tails.  This is because the residual $\Delta_{\mu}$ is raised to the power $i$, so an unbiased estimate of the moment depends on being able to sample the long tails of the distribution.  For sparse data, missing these long tails, the missing $\Delta^i_{\mu}$ has a large effect on the estimated moment.  

One approach to account for the bias introduced in sparse moments estimates is to use h-statistics \citep{halmos1946}, which correct the moments estimates based on the number $n$ of points in the sample.  For example, the corrected estimate $\hat{\mu}$ for the second moment is given by
\begin{equation}
\label{eq:h_stat_two}
\hat{\mu}_2= \frac{n}{n-1} \mu_2,
\end{equation}
where $\mu_2$ is given by equation (\ref{eq:moments_def_discrete}).  Similarly, the estimate of the third moment is given by
\begin{equation}
\label{eq:h_stat_three}
\hat{\mu}_3= \frac{n^2}{(n-1)(n-2)} \mu_3.
\end{equation}
Note that in the limit of large $n$, these estimates converge to the unweighted estimates.  However, we find in tests on simulated realizations that the h-statistic estimator is sensitive to the sparsely sampled long tails of the probability distribution.

\subsection{Kernel Density Estimation}
\label{sec:kde}

\begin{figure}
\begin{center}
 \includegraphics[width=9cm]{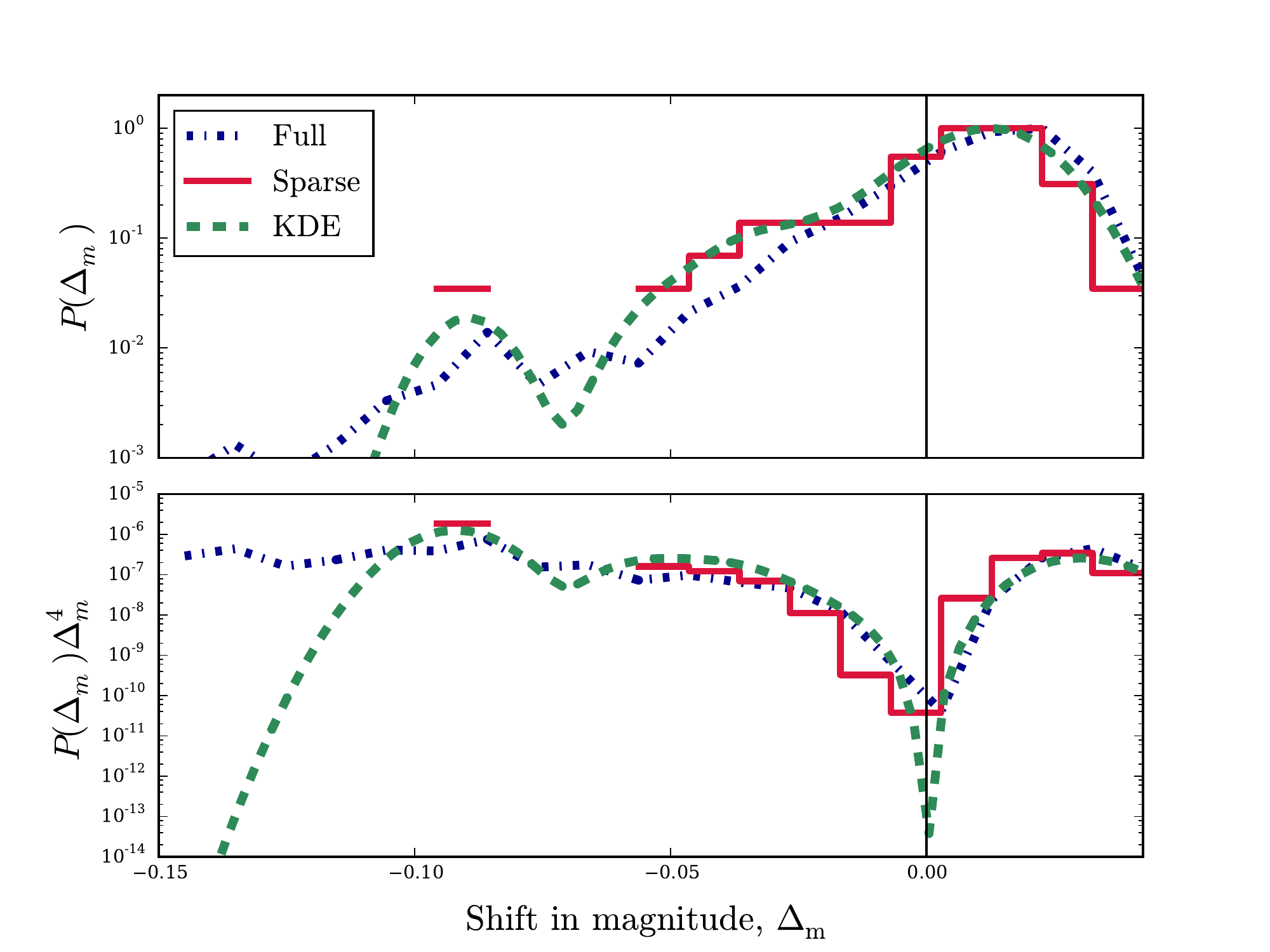}
\caption{Illustrating the KDE reconstruction of the magnitude residuals.  The top panel illustrates histograms of the lensing magnitude residuals between $0.6<z<0.8$.  The blue dash-dotted line (`Full') shows a histogram  of the distribution of residuals of 5500 sources.  The red solid line (`Sparse') shows the same histogram, but down sampled to 100 sources (typical of the number of supernovae per bin with the JLA catalogue).  The green dashed line (`KDE') illustrates the PDF of the residuals estimated only from the sparse sample.  The high magnification tail of the KDE-reconstructed PDF is closer to the full sample than the sparse.  To illustrate the effects on the moments, in the lower panel, the same PDFs are plotted, multiplied by $\Delta_{\rm{m}}^4$.  Although the KDE method does not reconstruct the very highly magnified ($\Delta_{\rm{m}}<-0.1$) tail, it is closer to the full distribution than the sparse sample.}
   \label{fig:KDE_example}
\end{center}
\end{figure}

To improve our estimates of the moments, we use Kernel density estimation (KDE) to provide a better estimate of the genuine underlying probability density function (PDF) of the magnitude residuals \citep[e.g.][]{1986desd.book.....S,2004ApJS..155..257R}.  The KDE method we use here convolves each point in the sample of residuals with a normal distribution.  The bandwidth $h$ of the normal distribution is determined by `Scott's Rule' \citep{WICS:WICS103}, which, for one-dimensional data is given by
\begin{equation}
\label{eq:scott}
h = n^{-1/5} \sigma,
\end{equation}
where $n$ is the number of data points and $\sigma$ is the standard deviation of the data.  We then use equation (\ref{eq:moments_def_integral}) to calculate the moments, with the probability distribution estimated with KDE.  The reconstruction is illustrated in Fig. \ref{fig:KDE_example}, where we use KDE to estimate a distribution of residuals from a random sparse sub-sample of a full population.  The full sample contains 5500 measurements, from which we have randomly sampled 100 measurements in the sparse sample.  

\begin{figure*}
\begin{center}
 \includegraphics[width=16cm]{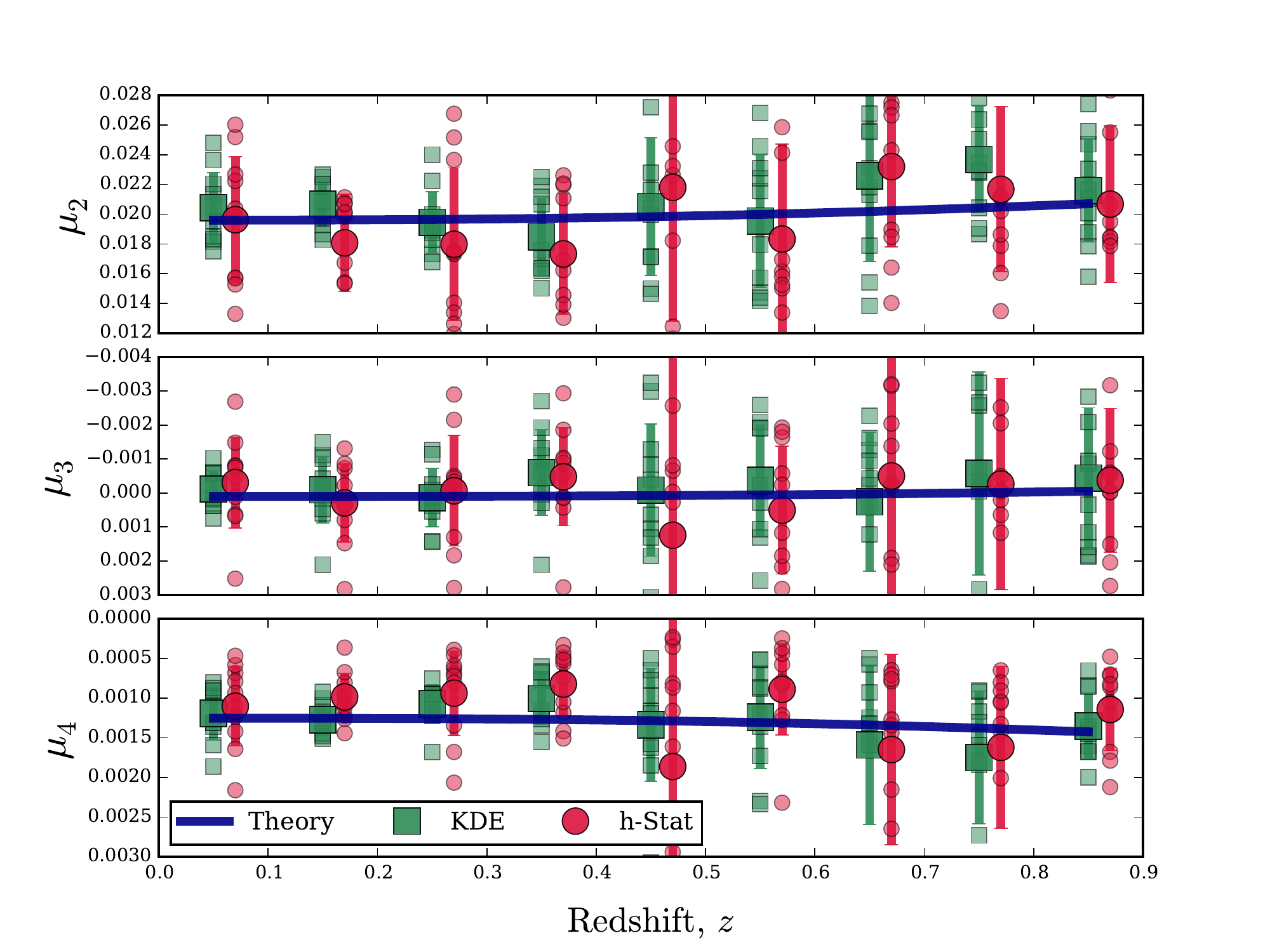}
\caption{Comparing methods for measuring the moments.  The red circles illustrate moments estimated with equation (\ref{eq:moments_def_discrete_weight}), a weighted summation of the residuals) and corrected using h-statistics, such as equations \ref{eq:h_stat_two} and \ref{eq:h_stat_three}.  The green squares illustrate the moments estimated with the KDE method.  The smaller points illustrate the measured moment for each of the 10 simulated realizations.  The larger points illustrate the ensemble average of these realizations. The error bars indicate the standard deviation of the measured moments.  We note that the KDE method is more accurate and precise than the h-statistic method -- the measurements are closer to the theoretical values and the scatter is smaller.}
   \label{fig:KDE_example_moments}
\end{center}
\end{figure*}

The effect of the KDE moments is shown in Fig. \ref{fig:KDE_example_moments}, where we compare the KDE estimate of the moments to the estimate with equation (\ref{eq:moments_def_discrete_weight}) with the h-statistic correction, using the simulated catalogues described in section \ref{sim_section}.  We note that the KDE method appears to be more precise and accurate than the standard moments estimators.

\subsection{Estimating the Covariance Matrix}
\label{sec:cov_mat}

In \cite{2014PhRvD..89b3009Q}, the covariance matrix was calculated from the analytic expectation for the covariance between the moments, given by equation 24 from \cite{2014PhRvD..89b3009Q}.  We find that this covariance matrix leads to a high $\chi^2$ per degree of freedom in our parameter fits.  This may be because this expression for the covariance matrix depends on all the moments up to the 8$^{\rm{th}}$, and as such is particularly sensitive to any outliers, the effects of which would be amplified when measuring such high moments of the distribution.

\begin{figure*}
\begin{center}
 \includegraphics[width=16cm]{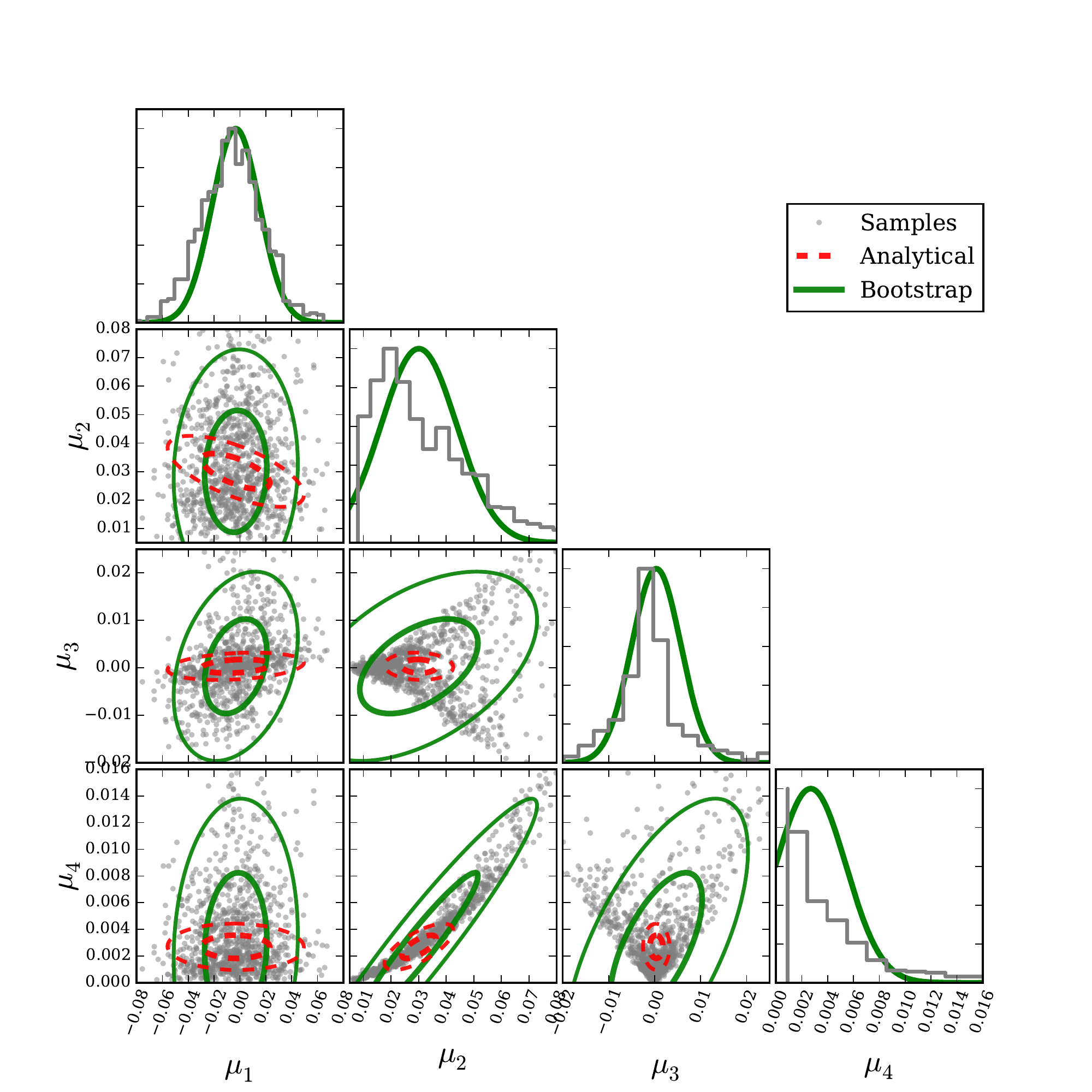}
\caption{Illustrating the bootstrap re-sampling method to estimate the covariance matrix.  The red dashed ellipses illustrate the covariance matrix estimated with equation 24 from \protect\cite{2014PhRvD..89b3009Q}.  The grey points illustrate bootstrap resamples of the JLA survey.  The green, solid ellipses illustrate the covariance matrix fitted to these bootstrap resamples.  The ellipses are plotted at the one and two standard deviation levels.  We note that both techniques depend only on the observed moments of the survey (and do not depend on the physical model for the dispersion).  The bootstrap method depends only on measuring moments up to the fourth, whereas the analytical method relies on measuring up to the eighth moment. }
   \label{fig:covariance_matrix}
\end{center}
\end{figure*}

Instead, our approach is to use a bootstrap re-sampling technique to estimate the covariance matrix. We randomly sub-sample a fraction of the survey, which we treat as a realization.   We repeat this sampling to generate multiple realizations, from which we can then directly measure the covariance matrix.  An example of the technique is shown in Fig. \ref{fig:covariance_matrix} for one redshift bin (in the range $0.4<z<0.5$).  The grey points show the measured moments for each of the samples.  The green, solid line, ellipses show the one and two standard deviation levels of the best-fitting covariance matrix to these samples.  The dashed red ellipses show the equivalent covariance matrix estimated with equation 24 from \cite{2014PhRvD..89b3009Q}. We note that both methods depend only on the observed moments, and not the theoretical modelling of lensing or velocities. 

We choose a sampling fraction of 70$\%$ to give a $\chi^2$ per degree of freedom close to one.  Increasing this fraction increases the $\chi^2$, since there is less variation between the samples.  We note that this method of estimating the covariance matrix gives the same uncertainty in $\mu_1$ as the analytical expectation for the term in the covariance matrix, given by $\mu_2/N_j$, where $N_j$ is the number of supernovae in the redshift bin.

\section{Results \& Discussion}
\label{sec:results}

In this section, we present results of fitting our velocity and lensing likelihood to simulated and genuine data.

\subsection{Tests with Simulated Data}

\begin{figure*}
\begin{center}
 \includegraphics[width=16cm]{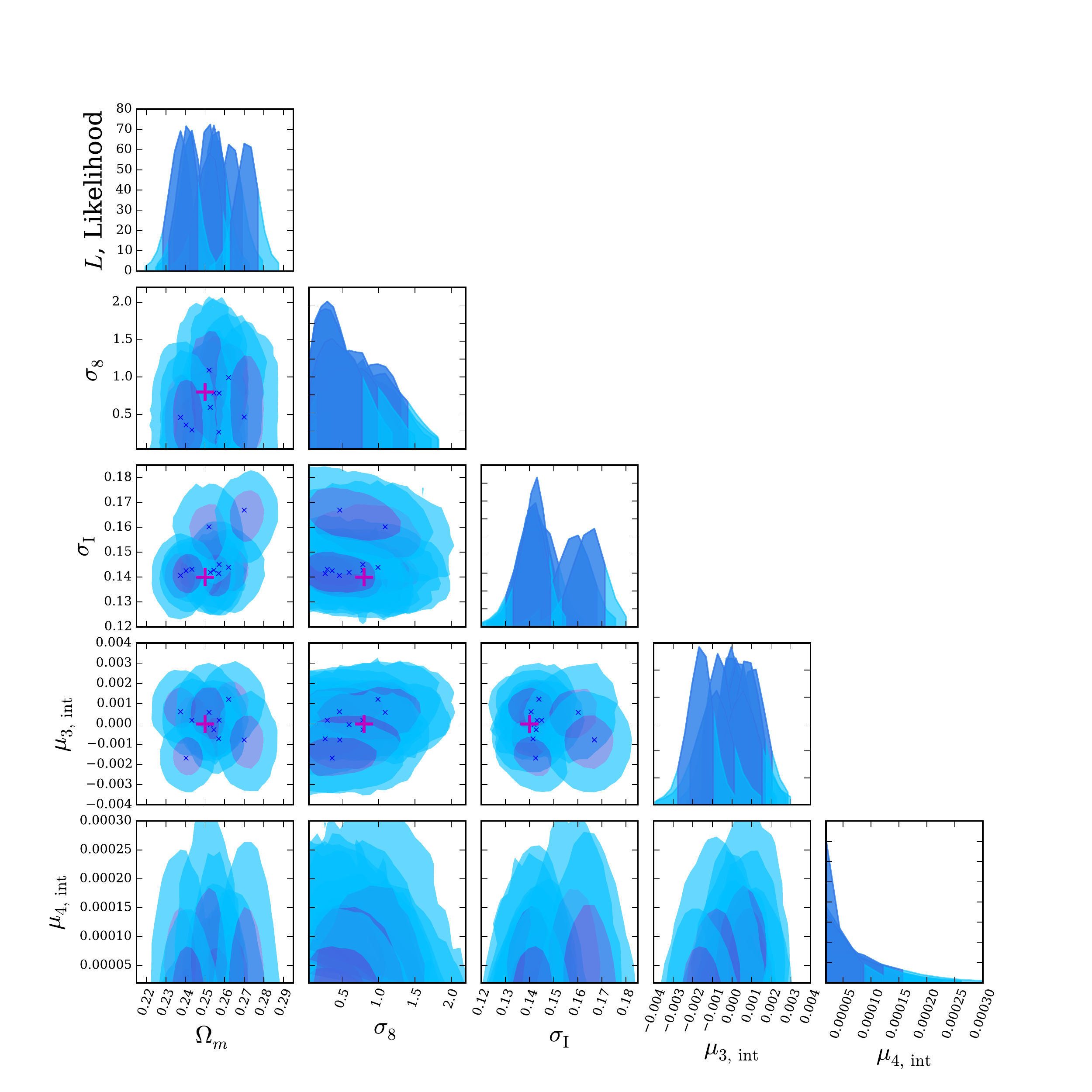}
\caption{Testing the moments likelihood with lensing and velocities on simulated realizations of the JLA survey from the MICE simulation.  The horizontal crosses indicate the input values in the simulated catalogues, and the diagonal crosses indicate the recovered values from the likelihood.  We note that when allowing the intrinsic supernova dispersion to vary (parametrized by the intrinsic dispersion $\sigma_{\rm{I}}$ and the intrinsic third and fourth moments), we find that the intrinsic dispersion is systematically over-estimated and $\sigma_8$ is systematically underestimated.}
   \label{fig:triangleSim}
\end{center}
\end{figure*}

Before applying our likelihood to the genuine JLA catalogue, we first test the likelihood on the simulated realizations of the JLA.  In Fig. \ref{fig:triangleSim} we illustrate the results of fits to the simulated realizations. 
We fit for $\Omega_{\rm{m}}$, $\sigma_8$, intrinsic dispersion, $\sigma_{\rm{I}}$, and intrinsic third and fourth moments, $\mu_{3, \, \rm{int}}$ and $\mu_{4, \, \rm{int}}$.  We find when fitting for this full parameter set, $\sigma_8$ is systematically underestimated, and $\sigma_{\rm{I}}$ is overestimated.  The average value of $\sigma_8$ recovered with this parametrization is $75 \%$ of the true simulation input value.  We believe that the main cause of this bias is a degeneracy between the intrinsic dispersion and the combined effects of lensing and velocities.

As can be seen in Fig. \ref{fig:moments_theory}, the combined effect of velocities and lensing on the second moment is almost independent of redshift across the range $0.1<z<0.5$ (the redshift range over which most of the JLA is distributed).  This is due to the dispersion from velocities decreasing with redshift, while the dispersion from lensing increases.  This cancellation of redshift dependence contributes to the degeneracy of the effects of $\sigma_8$ with the intrinsic dispersion.  The $\sim 0.01$ contribution to the second moment from $\sigma_8$ across this redshift range is consistent with the overestimate of the intrinsic dispersion in the simulated catalogues.  

\begin{figure}
\begin{center}
 \includegraphics[width=9cm]{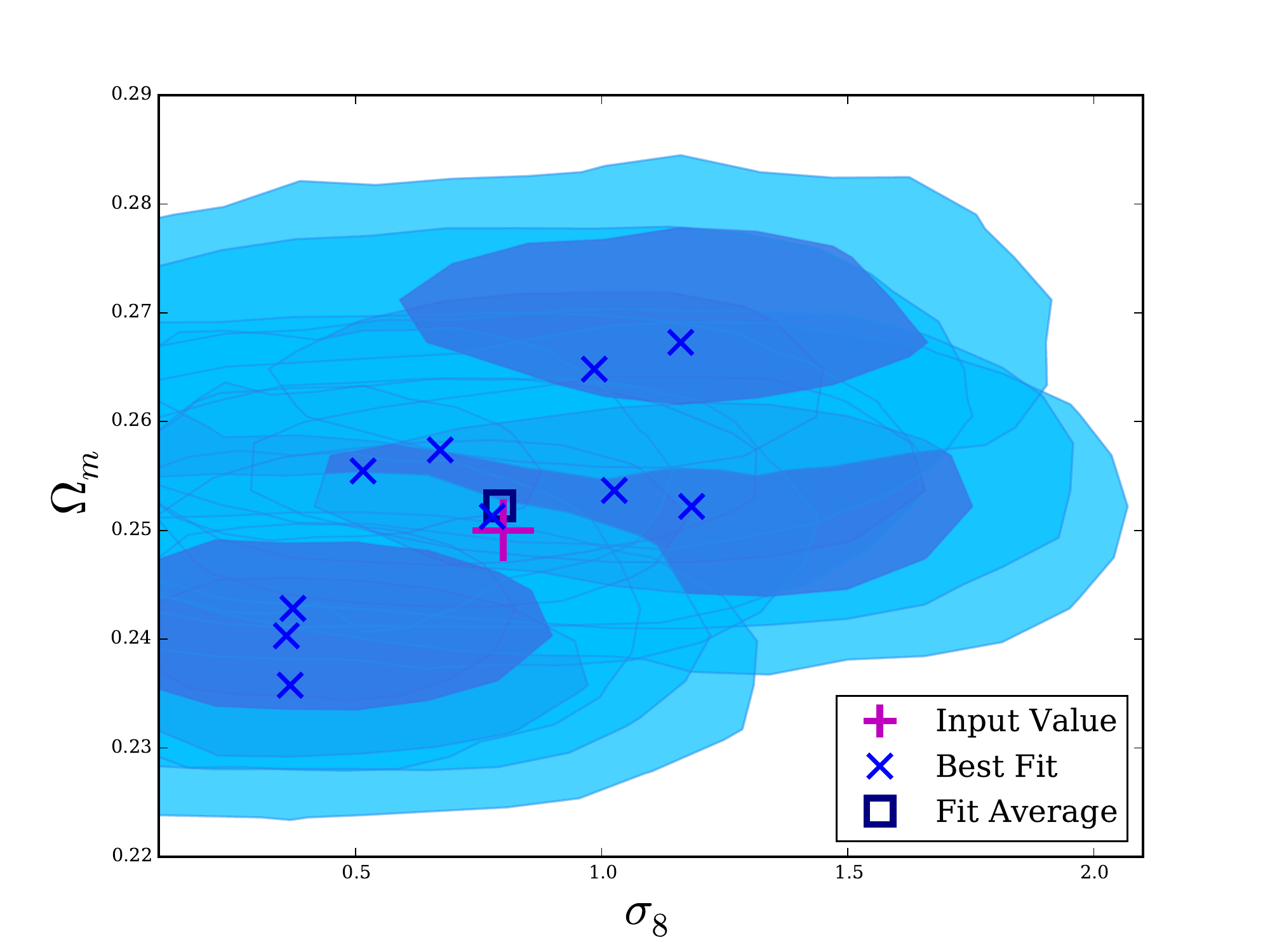}
\caption{Fitting simulated realizations of the JLA catalogue, with lensing and velocity effects.  Here, we are only fitting for $\Omega_{\rm{m}}$ and $\sigma_8$, and have kept the model for the intrinsic supernova dispersion fixed as a Gaussian distribution with dispersion $\sigma_{\rm{I}}=0.14$.  For each realization, we illustrate the best-fitting value with a diagonal cross, and the one and two standard deviation contours.  The dark blue square illustrates the average of these fits, and the magenta cross illustrates the simulation input value.}
   \label{fig:contour_sim}
\end{center}
\end{figure}

In Fig. \ref{fig:contour_sim}, we keep the intrinsic dispersion fixed at the input value of 0.14 mag, and the intrinsic third and fourth moments fixed at zero, varying only $\Omega_{\rm{m}}$ and  $\sigma_8$.  The average value of $\sigma_8$ recovered with this parametrization is 0.8, matching the input value of the simulation.  

\subsection{Results from JLA Data}

\begin{figure*}
\begin{center}
 \includegraphics[width=14cm]{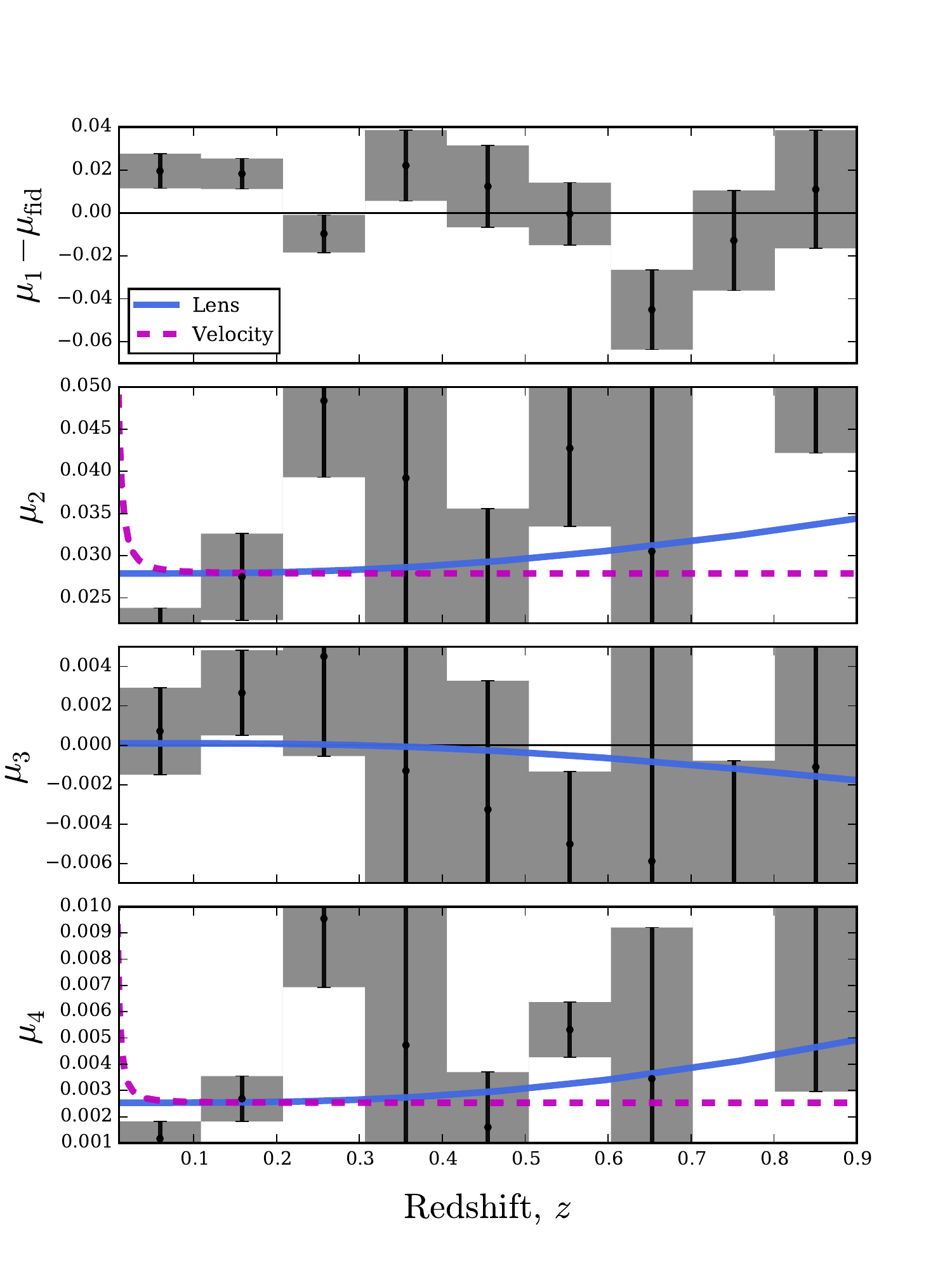}
\caption{Illustrating the moments measured from the magnitude residuals in the JLA catalogue.  The plotted uncertainties on the moments are the square-root of the diagonal of the moments covariance matrix.  The best-fitting lensing model is illustrated with the blue solid line, and the best-fitting velocity model is illustrated with the magenta dashed line.}
   \label{fig:Moments_measured_JLA}
\end{center}
\end{figure*}

\begin{figure*}
\begin{center}
 \includegraphics[width=18cm]{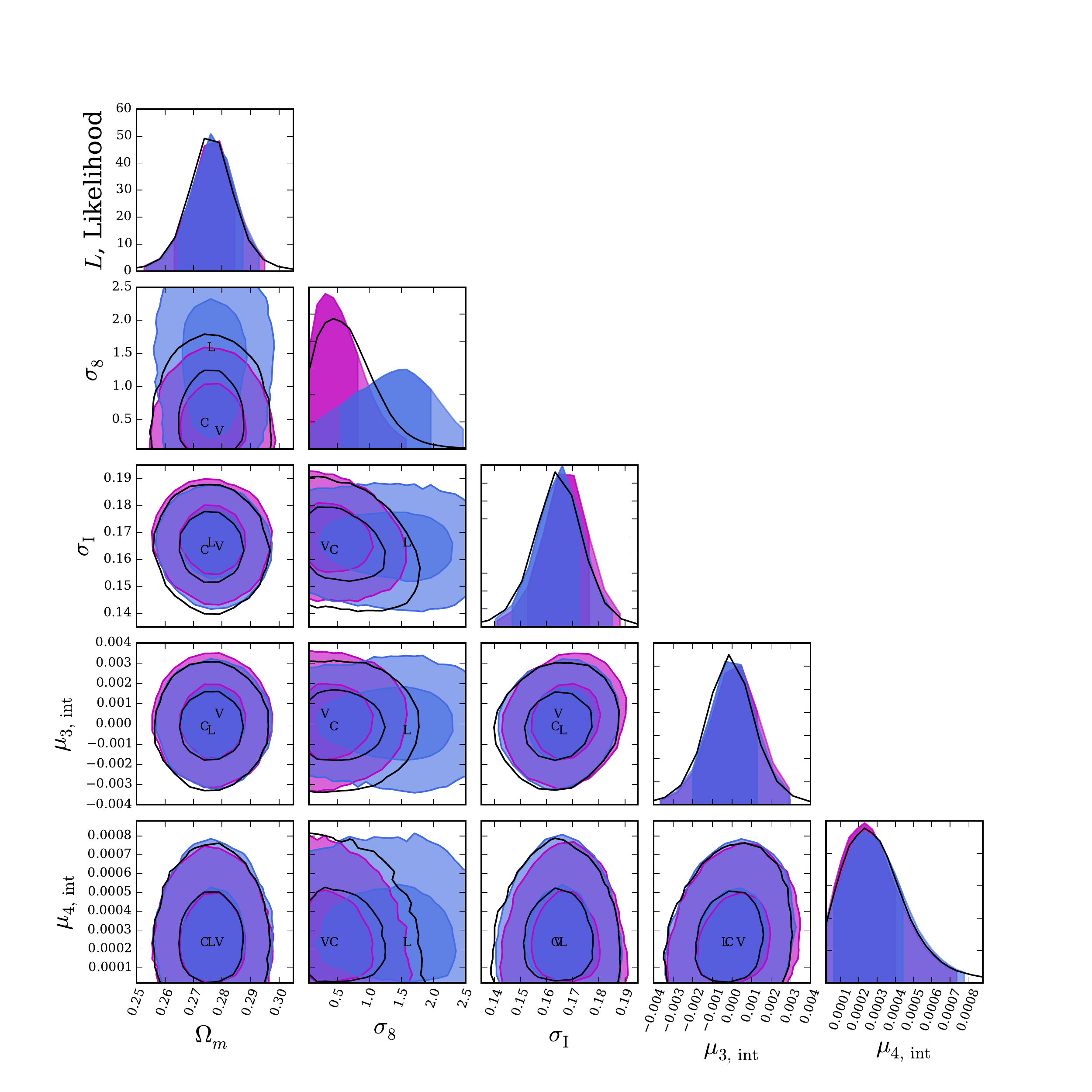}
\caption{Parameter constraints for the genuine JLA survey.  The best-fitting values for the lensing-only model are shown in blue with an `L' -- the best-fitting velocity model is shown in magenta with `V' and the combined results in black with `C'.  The combined result is closest to the velocity only model due to the greater number of low-redshift supernovae in JLA, where the effects of velocities are more significant than lensing.  We note that in tests on simulations, this method tends to overestimate the intrinsic dispersion, and underestimates the value of $\sigma_8$. }
   \label{fig:Triangle_JLA}
\end{center}
\end{figure*}

\begin{figure}
\begin{center}
 \includegraphics[width=9cm]{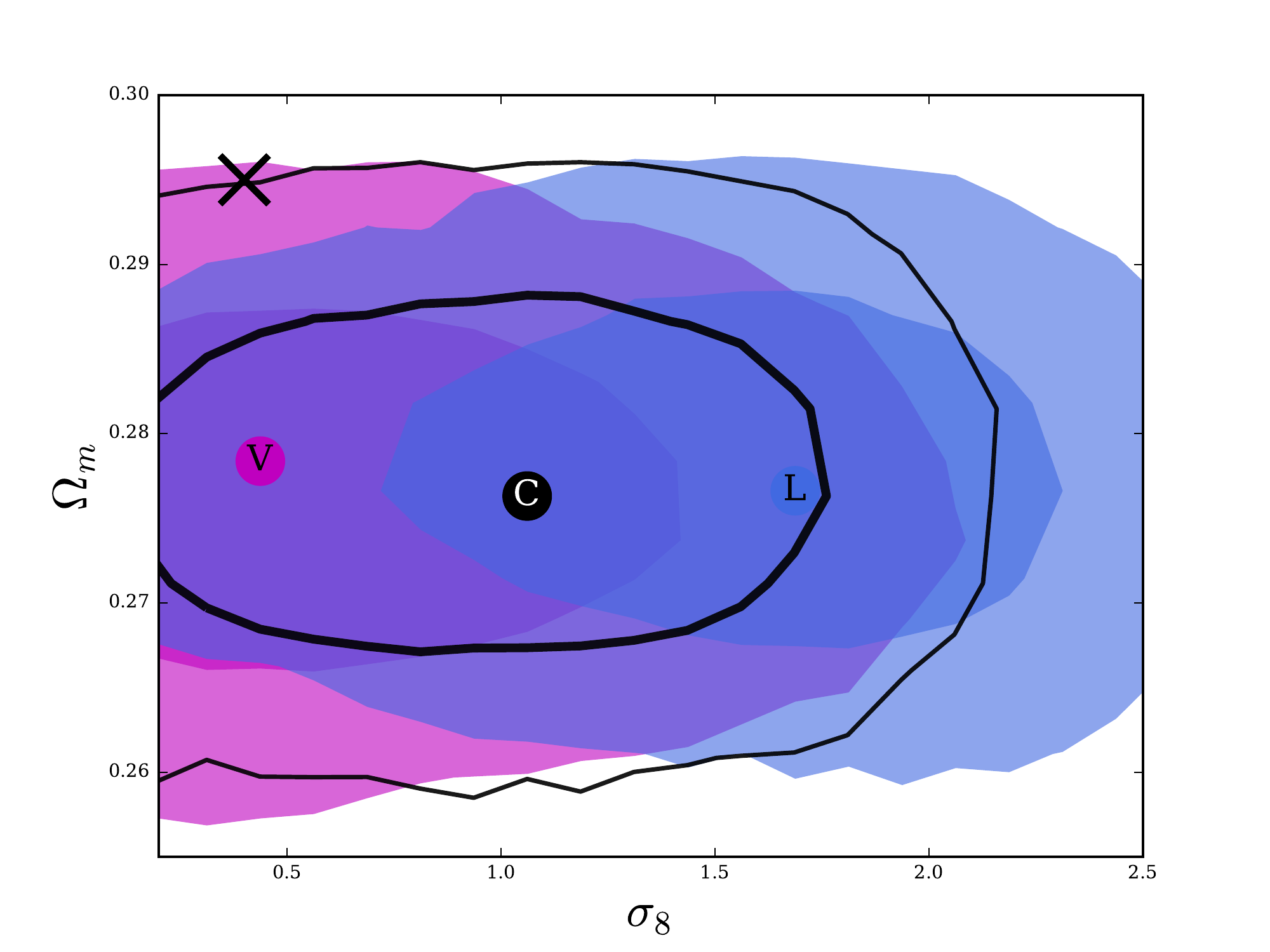}
\caption{Fitting the genuine JLA survey for $\Omega_{\rm{m}}$ and $\sigma_8$.  We have kept the model for the intrinsic supernova dispersion fixed as a Gaussian distribution with dispersion $\sigma_{\rm{I}}=0.14$.  The results for the lensing only fit are shown in blue, the velocity only fit in magenta, and the combined fit in black.  The cross indicates the value of $\Omega_{\rm{m}}$ from \protect\cite{2014A&A...568A..22B} and the value of $\sigma_8$ from \protect\cite{2016PDU....13...66C}.  }
   \label{fig:JLA_contour}
\end{center}
\end{figure}

We now apply the analysis method to the genuine JLA catalogue.  In Fig. \ref{fig:Moments_measured_JLA} we illustrate the moments measured with the KDE method. 
We use equation (\ref{eq:MeMO_like}) to fit the models for the moments to these measurements. We allow the intrinsic dispersion and moments to vary.  The results of the fits are summarized in Table \ref{tab:params} and illustrated in Fig. \ref{fig:Triangle_JLA}. 
\begin{table*}
  \centering
  \begin{tabular}{r c c c c c c}
& $\Omega_{\rm{m}}$ & $\sigma_8$ & $\sigma_{\rm{I}}$ & $\mu_{3, \,  \rm{int}}$ ($ \times 10^{-3} $)  & $\mu_{4, \, \rm{int}}$ ($ \times 10^{-4} $) & $\chi^2 / $ DoF  \\
\hline
Velocities & 0.279$\pm$0.013  &  0.32$^{+0.63}_{-0.32}$ &   0.17$\pm$0.02 &  0.5$\pm$2.0   &  2$\pm$2   &  1.14 \\ 
 \\ 
Lensing & 0.276$\pm$0.016  &   1.56$^{+0.51}_{-1.01}$   & 0.17$\pm$0.02  &  0.4$\pm$2.0 &  2$\pm$2  &   1.20  \\ 
 \\ 
Combined & 0.274$\pm$0.013  &  0.44$^{+0.63}_{-0.44}$    & 0.16$\pm$0.02 & -0.2$\pm$2.0  &  2$\pm$2    &  1.14   \\ 
 \\ 
  \end{tabular}
  \caption{Results of the fits for the lensing and velocity models, including fits to intrinsic dispersion parameters, all 68\% confidence intervals. }
  \label{tab:params}
\end{table*}
We also repeat the parameter estimation with the intrinsic dispersion fixed as a Gaussian distribution of width 0.14 mag; these results are summarized in Table \ref{tab:params_cosmo}, and illustrated in Fig. \ref{fig:JLA_contour}.  For both sets of parametrizations, we fit for the effects of velocities and lensing individually, and combined.
\begin{table}
  \centering
  \begin{tabular}{r c c c }
& $\Omega_{\rm{m}}$ & $\sigma_8$ & $\chi^2 / $ DoF  \\
\hline
Velocities & 0.275$\pm$0.012  &  0.44$^{+0.76}_{-0.44}$  &  1.26  \\ 
 \\ 
Lensing & 0.278$\pm$0.011  &    1.70$^{+0.51}_{-0.76}$    &  1.29 \\ 
 \\ 
Combined & 0.278$\pm$0.011  & 1.07$^{+0.50}_{-0.76}$  &  1.27   \\ 
 \\ 
  \end{tabular}
  \caption{Results of the fits for the lensing and velocity models, with intrinsic dispersion parameters fixed, all 68\% confidence intervals. }
  \label{tab:params_cosmo}
\end{table}
We find a value of $\Omega_{\rm{m}}=0.274 \pm 0.013$, which is lower than the value of $\Omega_{\rm{m}}=0.289\pm0.018$ from \cite{2014A&A...568A..22B} (with a flat $\Lambda$CDM model and quoting the statistical uncertainty).  We believe the main cause of this difference is due to the peculiar velocity corrections used by \cite{2014A&A...568A..22B}, which we do not use in our fits since we are modelling the effects of the velocities as a signal.  Repeating our fits without the peculiar velocity correction, we find $\Omega_{\rm{m}}=0.305\pm0.027$.  We also truncate our analysis for $z>0.9$, so our data set is not identical to \cite{2014A&A...568A..22B}.

Keeping the intrinsic dispersion fixed at 0.14 mag, we find values of $\sigma_8$ shown in Table \ref{tab:params_cosmo}:   $1.07^{+0.50}_{-0.76}$ for the combined model.  When we allow the intrinsic dispersion to vary (and also fit for intrinsic third and fourth moments) we find lower values for $\sigma_8$, shown in Table \ref{tab:params}.

We find values for the intrinsic third and fourth moments that are consistent with a purely Gaussian intrinsic supernova distribution.  Our value for the magnitude of the dispersion, $\sigma_{\rm{I}}=0.162\pm0.016$ is higher than the value of 0.14 found by \cite{2014MNRAS.443L...6C}.  We note that in tests on simulations, our method overestimates the value of $\sigma_{\rm{I}}$ by $\sim0.01$ mag, and as such this result is unlikely to suggest a tension in the value of the intrinsic dispersion.  

With this parametrization (where we allow the model for the intrinsic dispersion to vary), we find a value of $\sigma_8$ of $0.44^{+0.63}_{-0.44}$ (although we note that in the simulations, this parametrization underestimates the value of $\sigma_8$ by $\sim 25\%$).  For comparison, \cite{2016PDU....13...66C} found $\sigma_8 = 0.40^{+0.21}_{-0.23}$ with a different approach to combining the effects of lensing and velocities, using two separate likelihoods for each effect (with the growth index $\gamma$ fixed at the expected value in GR of 0.55).  The larger uncertainty in our value of $\sigma_8$ is consistent with the larger uncertainty estimates of the data covariance matrix from the bootstrap re-sampling approach we use, as illustrated in the comparison of covariance matrices in Fig. \ref{fig:covariance_matrix}.

\begin{figure}
\begin{center}
 \includegraphics[width=9cm]{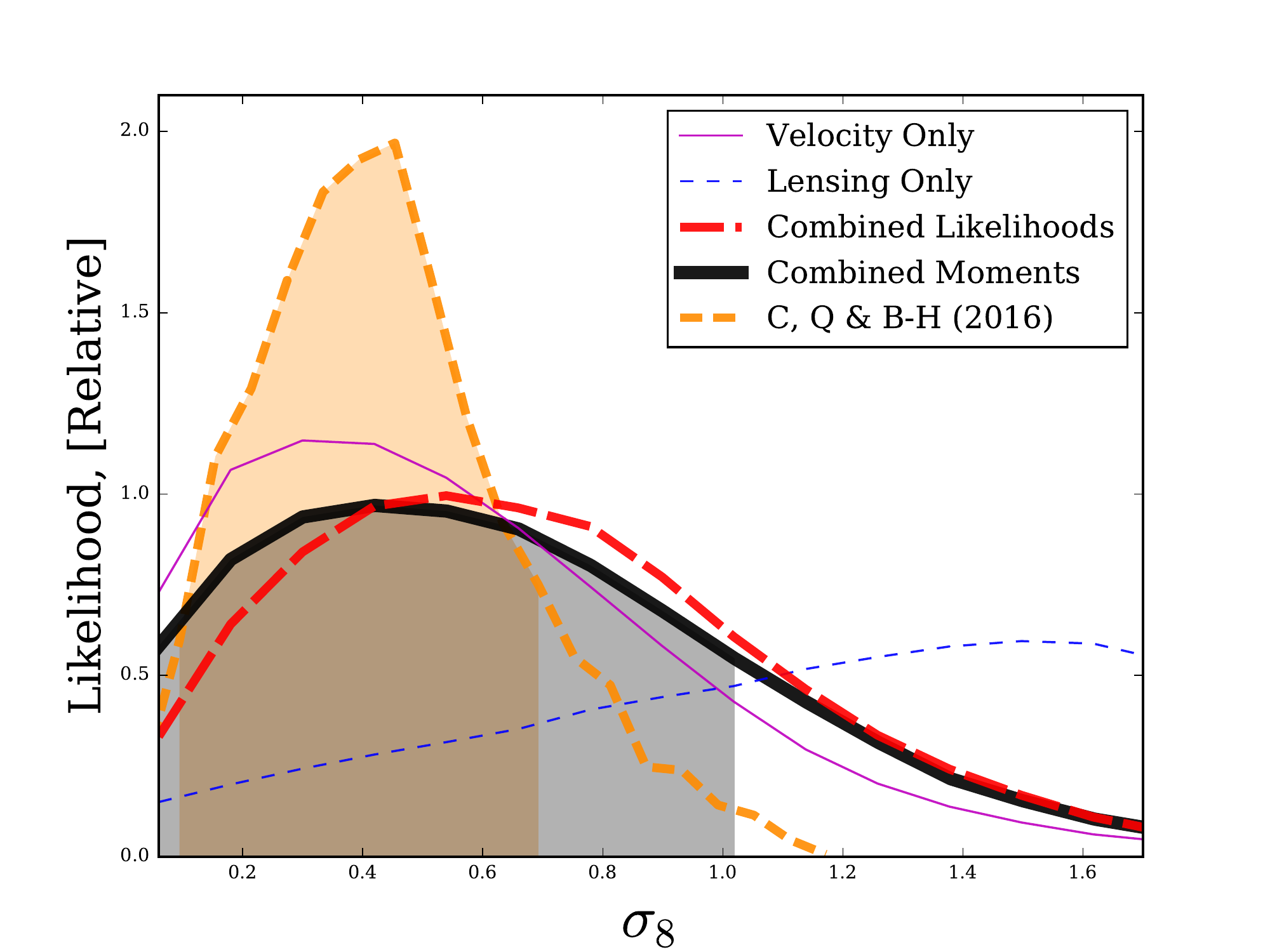}
\caption{Comparing marginalized values of $\sigma_8$.  The thin, solid magenta line, labeled `Velocity Only' illustrates the marginalized likelihood on $\sigma_8$ fitting only for velocities.  The thin, dashed blue line shows the same constraint when fitting only for lensing.  The thick, solid black line, labeled `Combined Moments' shows the result from fitting simultaneously to these effects in our model for the moments.  These three marginalized likelihoods are shown in full in Fig. \ref{fig:Triangle_JLA}.  We compare our `Combined Moments' fit to the `Combined Likelihoods' result, plotted with a dashed red line, which is the product of the velocity and lensing likelihoods (normalized to unit area).  We also compare our results to \protect\cite{2016PDU....13...66C}.  }
   \label{fig:sig_8_marg_compare}
\end{center}
\end{figure}

We believe that the approach of \cite{2016PDU....13...66C} (splitting the supernovae into high and low redshift subsets) yields a similar value of $\sigma_8$ to our approach because both results are driven primarily by the low-redshift peculiar velocity results, due to the greater number of supernovae at low redshift.  We can see in Fig. \ref{fig:sig_8_marg_compare} that our marginalized constraint on $\sigma_8$ with the combined moments and velocity model (shown with a thick, solid black line) is closer to the velocity-only model (shown with a thin, solid magenta line) than to the lensing-only model (shown with a thin, dashed blue line).  We note that the product of these two individual likelihoods (illustrated with a thick dashed red line) is similar to the result from the combined moments -- the addition of the lensing model has only a small effect on the final values.  We also illustrate the marginalized likelihood from \cite{2016PDU....13...66C} \footnote{We have downloaded the MCMC chains from \textsc{http://sites.if.ufrj.br/castro/en/pesquisa/artigos/}}.  We note that these values of $\sigma_8$ are consistent with the low amplitude of the peculiar velocity covariance matrix in the JLA analysis of \cite{2015JCAP...12..033H}.  However, we note that the velocity and lensing effects are of a comparable magnitude over the redshift range $0.2<z<0.5$, and, as such, it will be important to model these effects simultaneously for future, deeper supernova surveys.

Our lensing-only result is higher than that found by \cite{2014MNRAS.443L...6C}, who found $\sigma_8=0.84^{+0.28}_{-0.65}$ at the 68\% confidence level, or $\sigma_8 < 1.45$ at the 95\% confidence level.  We note that using the moments estimator in equation (\ref{eq:moments_def_discrete}), and the covariance matrix from equation 24 from \cite{2014PhRvD..89b3009Q}, we can reproduce the value of $\sigma_8$ from \cite{2014MNRAS.443L...6C}, finding $\sigma_8=0.89^{+0.35}_{-0.59}$ (68\% confidence level), or $\sigma_8 < 1.60$ at the 95\% confidence level.  We also find an intrinsic dispersion of $\sigma_{\rm{I}}$ = 0.14$\pm$0.01, and intrinsic third and fourth moments that are consistent with zero.  However, we note that this method is more biased towards underestimates than the KDE method.  We also note that \cite{2014MNRAS.443L...6C}  use corrections to this estimator to account for bias due to the sparsity of the samples, although the details of the corrections are not published.

We now consider some effects that may lead to a higher value for $\sigma_8$ from lensing.  Our analysis has assumed that both the large-scale ($k<0.1\, h$ Mpc$^{-1}$) density fluctuations relevant to peculiar velocities and the small-scale ($k>1.0 \, h$ Mpc$^{-1}$) fluctuations relevant to lensing can be set by $\sigma_8$.  Taking the results at face value, we might interpret the lensing and velocity results as suggestive of a tilt in the matter power spectrum, suppressing large scale power and boosting small scale power.  The small scale power spectrum in particular is sensitive to baryonic feedback, which remains challenging for theory and simulation \citep[e.g.][]{2012MNRAS.426..140D,2012MNRAS.419..465D,2012MNRAS.423.1596S,2013MNRAS.429.1922C}.

\cite{2014A&A...568A..22B} note that due to Malmquist bias, the intrinsic dispersion of the highest redshift supernova magnitudes may decrease by 0.01 mag (for comparison, at $z=1$, the dispersion due to lensing for $\sigma_8=0.8$ is 0.04 mag).  In our model of a constant intrinsic dispersion, this would lead to an overestimate of the lensing dispersion to compensate.  For example, at $z=1$, $\sigma_8=0.8$, and an intrinsic dispersion of 0.150 mag, the combined dispersion is 0.155 mag.  To get the same total dispersion with an intrinsic dispersion 0.01 mag lower, we require a value of $\sigma_8=1.3$.

\section{Conclusions}
\label{sec:conclusions}

We have considered the effects of peculiar velocities and lensing magnification on moments of the distance redshift relation.  We have described theoretical models for these effects, and statistical methods for undertaking these measurements on sparse data.  

We note that at redshift $z\sim0.2$--$0.5$, the effects of peculiar velocities and lensing magnification contribute similarly to the dispersion in the Hubble diagram. The cancellation of redshift dependence of these two effects across this redshift range makes the effects of $\sigma_8$ on the second moment in this redshift range degenerate with the intrinsic supernova dispersion.  We thus emphasize the importance of modelling both effects simultaneously, and also the importance of measuring higher moments of the Hubble diagram, in order to break this degeneracy.

We confirm that the simulated lensing convergence in the MICE light cone simulation is in excellent agreement with the moments modelled by  \cite{2013PhRvD..88f3004M}.  We present an extension of the MeMo likelihood of \cite{2014PhRvD..89b3009Q}, which directly includes the effects of peculiar velocities in a single likelihood.  We note that standard estimators for the moments of the magnitude residuals underestimate the genuine moments for typical numbers of supernovae.  We show that KDE can be used to reduce the bias in estimates of the moments from sparse samples.  We then apply the velocity and lensing likelihood and the KDE estimators to the JLA supernovae catalogue.  

Comparing to other work, we note that using the moments estimator of equation (\ref{eq:moments_def_discrete}), we can reproduce the values of $\sigma_8$ and $\sigma_{\rm{I}}$ from \cite{2014MNRAS.443L...6C}.  However, our result with this estimator is likely to be an underestimate of the genuine moments of the JLA survey.

There are some limitations to the current analysis: as with other work focusing on the Hubble diagram residuals \citep{2014ApJ...780...24S,2014MNRAS.443L...6C}, we do not account for correlations in the distance moduli in our analysis (either from the light-curve fitting, or density perturbations).  This approach frees us from the imposition of Gaussianity (which is implicit in a typical covariance-matrix analysis) in the distribution of residuals, which is the dominant signal for lensing, and the primary focus of this paper.  In the case of lensing, since the signal depends on the integrated line of sight density, we expect the magnification effect to be uncorrelated for angular separations greater than a few arcminutes. However, we note that \cite{2016arXiv161101315S} found that the magnitude residuals in the JLA catalogue are consistent with zero correlations from lensing magnification.  

In the case of peculiar velocities, we do expect large-scale correlations, although we have verified by calculating the full covariance matrix that the correlations are negligible for all but the lowest redshift supernovae.  In the case of the light curve parameters, \cite{2016PDU....13...66C} showed that marginalizing over the values slightly increased the uncertainties, but did not significantly bias the results or conclusions.  

Currently, the main limitation in the analysis is measuring unbiased moments of sparse samples of the residuals.  However, as the size of supernova catalogues increases, this issue will become less problematic.  With larger catalogues, however, it will become more important to model the intrinsic dispersion in the supernovae, such as the dependence with redshift and correlations with host galaxy type.  We emphasize that it is possible to place limits on the amplitude of the matter power spectrum with the supernovae Hubble diagram -- both the background expansion and perturbations with the same observable.  By fitting for the effects of lensing and velocities, we can also test for consistency in the Newtonian and lensing potentials.  Furthermore, by measuring the moments of the residuals, we test for consistency or evidence for outliers in the supernova population.

\section{Acknowledgments}

We thank Miguel Quartin for advice about the MeMo covariance matrix, Douglas Scott and Chris Blake for useful comments and suggestions on a draft of this paper, Michael Bromley for assistance with computing, and Qiuyue Liang for useful comments.  This research was conducted by the Australian Research Council Centre of Excellence for All-sky Astrophysics (CAASTRO), through project number CE110001020.  DS and TC thank Faculty of Technology at the University of Portsmouth for their support, and DB and RCN acknowledge funding from STFC grant ST/N000668/1.

\bibliographystyle{mn2e}
\bibliography{SupernovaeSigma8}

\end{document}